\documentclass[prl,reprint,twocolumn]{revtex4}
\usepackage{times}
\usepackage{amsmath,graphicx}

\newcommand{\Hsq}{\tilde{H}}

\begin{document}

\title{Mixed quantal-semiquantal dynamics with stochastic particles for backreaction} 

\author{Koji Ando\footnote{E-mail: ando@kuchem.kyoto-u.ac.jp}}

\affiliation{
Department of Chemistry, Graduate School of Science, Kyoto University,
Sakyo-ku, Kyoto 606-8502, Japan}

\begin{abstract}
A mixed quantal-semiquantal theory is presented in which
the semiquantal squeezed-state wave packet describes the heavy degrees of freedom.
We first derive mean-field
equations of motion from the time-dependent variational principle.
Then, in order to take into account the interparticle correlation, in particular 
the \lq quantum backreaction\rq\ beyond the mean-field approximation,
we introduce the stochastic particle description for 
both the quantal and semiquantal parts. 
A numerical application on a model of O$_2$ scattering from a Pt surface demonstrates
that the proposed scheme gives correct asymptotic behavior of the scattering probability,
with improvement over the mixed quantum-classical scheme with Bohmian particles,
which is comprehended by comparing the Bohmian and the stochastic trajectories.
\end{abstract}

\maketitle

Mixed quantum-classical (MQC) dynamics have been a subject of interest 
not only in 
chemical physics \cite{Pechukas69,Meyer79,Bittner95,Micha96,Martens97,Sholl98,Kapral99,Thoss99,Gindensperger00,Prezhdo01,Deumens01,AndoSanter03,Burghardt04}
but also in quantum gravity \cite{Anderson95,Halliwell98}, 
cosmology \cite{Hu95b,Campos97}, 
and measurement \cite{Machida80,Zurek03}.
One major problem lies in the description of correlation
between the two parts, in particular,
the force from the delocalized quantal part to the localized classical part,
that is, the problem of \lq quantum backreaction\rq. 
It is intimately related to the description of non-adiabatic transitions
in which the Born-Oppenheimer approximation breaks down,
for instance, near the conical intersections of adiabatic states.
Many theories have been proposed,
but the problem is inherently of approximate nature \cite{Terno06,Salcedo12}.
Thus, the assessment would be based not only on the theoretical consistency 
but also on the practical accuracy in applications.
In addition, simplicity for computational implementation to realistic
systems will be an important aspect.

In chemical physics, 
the quantum part usually represents electrons or protons,
and the classical part represents heavier nuclei.
For the latter, localized wave packet (WP) description,
typically by Gaussian WPs \cite{Heller76,Littlejohn86b}, is also useful.
In recent years, we have been studying a
\lq semiquantal\rq\ (SQ) squeezed-state WP theory for
chemical problems, with applications to 
hydrogen-bond structure and dynamics \cite{Ando04hext,Ando05,Ando06,Sakumichi08,Kim09,Kim10,Ono12,Ono13}. 
An extension to electron WPs with the valence-bond spin-couplings was also examined \cite{Ando09,Ando12},
and a combination of nuclear and electron WPs was applied to liquid hydrogen \cite{Kim12,Kim14}.
Following these, we put forward in this Letter a mixed quantal-semiquantal (MQSQ) theory.

We start with a
trial wave function and derive the equations of motion (EOM) by the
time-dependent variational principle.
The resulting EOM for the SQ part have the canonical Hamiltonian form for
the center and width variables of the WP.
The quantal part follows a time-dependent Schr{\"o}dinger equation (TDSE),
in which the potential energy function is averaged over the SQ WP and thus includes
the WP variables as the time-dependent external parameters.
The potential function for the evolution of the SQ part is an average over both
the SQ WP and quantal wave function,
and thus we encounter the problem of \lq backreaction\rq.
To address this, 
we propose in this work to exploit 
the theory of stochastic particle (SP) dynamics \cite{Nelson66,Yasue81}.
The SP dynamics are described by
the stochastic differential equations (SDE)
whose Fokker-Planck form is
equivalent to the 
TDSE.
We thus describe both the quantal and SQ wavefunctions by the corresponding sets of SPs.
By assuming the pre-averaged form for the interaction between the SPs,
the interparticle correlation beyond the mean-field approximation
is described.

The coordinates of the quantal and SQ parts are represented
by $x$ and $X$.
For simplicity, we consider the SQ WP
of the form \cite{Arickx86,Tsue92,Pattanayak94}
%\begin{equation}
\begin{multline}
\chi _{\Gamma} (X, t) 
=
N_t
\exp \left[
- \left( \frac{1}{4 \rho _t^2}
- \frac{i}{\hbar} \frac{\Pi _t}{2 \rho _t} \right)
\left( X - Q_t \right)^2 \right.
\\
\left.
+ \frac{i}{\hbar} P_t 
\left( X - Q_t \right)
\right] ,
\label{eq:SQWP}
%\end{equation}
\end{multline}
in which
$N_t = 1/( 2 \pi \rho _t ^2 )^{1/4}$.
The WP is characterized by
a set of time-dependent variables 
$\Gamma _t \equiv \{ Q_t, P_t, \rho _t, \Pi _t \}$,
where $Q_t$ and $\rho _t$
describe the WP center and width, 
$P_t$ and $\Pi _t$ are their corresponding conjugate momenta.
Generalization to a correlated multi-dimensional WP,
in which the variables are vectors and matrices,
has been implemented for a simulation of liquid water \cite{Ono12}, 
but the simpler form of Eq. (\ref{eq:SQWP}) would be appropriate for
this first presentation.

For the total wave function, we set forth a factorized form
\begin{equation}
\psi _{\gamma \Gamma} \left(x, X, t \right)
= \chi _{\Gamma} (X, t)
\hspace*{0.2em}
\varphi _{\gamma \Gamma} \left( x, t \right) .
\label{eq:psi}
\end{equation}
The idea behind this factorization will be discussed 
below.
The subscript $\Gamma$ indicates the dependence on 
the variables that characterize the SQ WP
of Eq. (\ref{eq:SQWP}).
Similarly, $\gamma$ consists of a set of variables 
that characterize the quantal wave function $\varphi$;
in applications to the electronic wave function,
they can be the coefficients of molecular orbitals 
or configuration interaction,
the Thouless parameters for Slater determinant,
or the electron WP variables. 
In some cases,
$\varphi$ 
may also depend parametrically on the 
SQ WP variables $\Gamma$,
as indicated by the subscript to $\varphi$ in Eq. (\ref{eq:psi}).
Recently, exact factorization of molecular wave functions to electronic and nuclear
parts has been discussed \cite{Abedi10,Cederbaum13}.
The idea here is rather simple;
as we will take into account the interparticle correlation
via the combination with the SP description,
we start with the factorized form Eq. (\ref{eq:psi}) 
in a sense to avoid double-counting of the correlation.

The time-dependence of the wave function $\psi_{\Gamma \gamma}$ is described by the
variables $\Gamma_t$ and $\gamma_t$ whose EOM are derived from
the time-dependent variational principle with
the action integral 
\(
{\cal S} = \int_{t_1}^{t_2} dt
\langle \psi (t) | 
i \hbar 
\partial_t
- \hat{H} 
| \psi (t) \rangle ,
\)
in which 
\begin{equation}
\hat{H} = \hat{T}_{\!x} + \hat{T}_{\!X} + v(x, X) 
\label{eq:hatHmol}
\end{equation}
is the Hamiltonian 
with the kinetic energies 
$\hat{T}_{\!x}$ and $\hat{T}_{\!X}$
and the potential energy $v(x,X)$.
With the trial wave function of Eq. (\ref{eq:psi}), 
the stationary condition
of the action ${\cal S}$ with respect to the variation of $\varphi$,
$\delta {\cal S} / \delta \varphi = 0$,
gives 
\begin{equation}
i \hbar \frac{\partial }{\partial t} 
\varphi _{\gamma \Gamma} \left( x, t \right)
= \left(
\hat{T}_{\!x}
+ V \left( x ; \Gamma _t \right)
\right)
\varphi _{\gamma \Gamma} \left( x, t \right) ,
\label{eq:TDSE}
\end{equation}
in which $V$ is the averaged potential over the SQ WP $\chi$,
\begin{equation}
V \left( x ; \Gamma _t \right)
= \int dX \left| \chi _{\Gamma} (X, t) \right|^2 v(x, X) .
\label{eq:barV}
\end{equation}
Equation (\ref{eq:TDSE}) has a form of
TDSE
affected by the external time-dependent variables $\Gamma _t$
that represent the SQ WP.
The variation with respect to the variables in $\chi$,
$\delta {\cal S} / \delta \Gamma = 0$,
gives the EOM of the canonical Hamilton form 
\begin{equation}
\dot{Q} =   \frac{\partial \Hsq}{\partial P},
\hspace*{0.5em}
\dot{P} = - \frac{\partial \Hsq}{\partial Q},
\hspace*{0.5em}
\dot{\rho} =   \frac{\partial \Hsq}{\partial \Pi},
\hspace*{0.5em}
\dot{\Pi} = - \frac{\partial \Hsq}{\partial \rho},
\label{eq:canonicalEOM}
\end{equation}
with the Hamiltonian in the extended phase-space $\Gamma$,
\begin{equation}
\Hsq
= 
  \frac{P^2}{2M} + \frac{\Pi^2}{2 M}
+ \frac{\hbar^2}{8 M \rho^2}
+ U_{\gamma} ( \Gamma ) ,
\label{eq:Hext}
\end{equation}
in which $M$ is the mass for $X$ and
\begin{equation}
U_{\gamma} ( \Gamma )
= \int dx \left| \varphi _{\gamma \Gamma} ( x ) \right|^2 V (x ; \Gamma) .
\label{eq:barbarV}
\end{equation}

In Eq. (\ref{eq:barV}),
the SQ coordinate $X$ is integrated to give $V (x ; \Gamma)$,
whereas in Eq. (\ref{eq:barbarV}), 
both $x$ and $X$ are
integrated to give $U_{\gamma} (\Gamma)$.
Therefore, the dynamics of quantal and SQ parts 
that follow Eqs. (\ref{eq:TDSE})--(\ref{eq:Hext}) 
are under the mutual \lq mean-field\rq,
which causes the problem of describing the \lq backreaction\rq.
To address this,
we propose in this work to deploy the theory of SP dynamics \cite{Nelson66,Yasue81}.
The SP dynamics 
are described by the SDE,
\begin{equation}
dx _t = \frac{\hbar}{m} \left( \nabla_{\!x} S + \nabla_{\!x} R \right) dt
+ \sqrt{\frac{\hbar}{m}} dW _t ,
\label{eq:SDEx}
\end{equation}
in which $m$ is the mass for $x$
and $W_t$ represent the standard Wiener process.
The SDE for the $X$ part has the analogous form.
The functions $R$ and $S$ are the real and imaginary parts of 
\(
\ln \psi (x, X, t) 
= R(x, X, t) + i S(x, X, t) .
\)
For the SQ WP of Eq. (\ref{eq:SQWP}), the SDE is
%\begin{equation}
\begin{multline}
dX_t =
\left[
\frac{P_t}{M}
+ \frac{\Pi _t}{M} \left( \frac{X_t - Q_t}{\rho _t} \right) 
\right.
\\
\left.
- \frac{\hbar}{2M \rho _t} \left( \frac{X_t - Q_t}{\rho _t} \right)
\right] dt
+ \sqrt{\frac{\hbar}{M}} dW_t .
\label{eq:SDEsq}
%\end{equation}
\end{multline}
The first two terms in the right-hand-side
correspond to the \lq current\rq\ velocity, 
whereas the third term is the \lq osmotic\rq\ velocity. 
The first term $P/M$ represents the ordinary velocity of the WP center.
The second term describes the breathing velocity of WP width, $\Pi /M$,
scaled by a factor $(X-Q)/\rho$, 
which indicates that the particles in the regions of WP tail move faster than those near the WP center.
The third term is also scaled by the same ratio $(X-Q)/\rho$, 
but has the opposite sign from the second term,
and the factor $\hbar/(2M\rho)$ implies its origin from the quantum uncertainty.

Equations (\ref{eq:SDEx}) and (\ref{eq:SDEsq}) 
gives the description equivalent to that of
the guide wave function $\psi (x,X,t)$.
Hence, as long as we employ the original
Eqs. (\ref{eq:TDSE})--(\ref{eq:barbarV}), 
the SPs will still be under the mutual mean-field.
Now we propose to replace 
$V (x ; \Gamma )$ 
in Eq. (\ref{eq:TDSE}) 
by the bare $v (x, X)$,
and $U_{\gamma} (\Gamma )$ 
in Eq. (\ref{eq:Hext})
by $V (x ; \Gamma )$,
in an aim to take into account the interparticle correlations.
Therefore, the calculation proceeds as follows.
(i) We introduce a set of SP pairs $\{ (x_{\alpha}, X_{\alpha}) \}$,
$\alpha = 1, 2, \cdots, N_{\rm sp}$,
distributed according to the initial wave function $\psi (x,X,0)$.
Each pair $(x_{\alpha},X_{\alpha})$
associates guide wave functions $\varphi_{\alpha}$ and $\chi_{\alpha}$.
(ii) We propagate them 
by 
\begin{equation}
i \hbar \frac{\partial }{\partial t} 
\varphi _{\alpha } \left( x, t \right)
= \left(
\hat{T}_{\!x}
+ v \left( x , X_{\alpha} \right)
\right)
\varphi _{\alpha } \left( x, t \right) ,
\label{eq:TDSESP}
\end{equation}
and
Eq. (\ref{eq:canonicalEOM}) with
\begin{equation}
\Hsq _{\alpha} =
  \frac{P_{\alpha}^2}{2M} + \frac{\Pi_{\alpha}^2}{2 M}
+ \frac{\hbar^2}{8 M \rho_{\alpha}^2}
+ V( x_{\alpha}; \Gamma_{\alpha} ) ,
\label{eq:HextSP}
\end{equation}
and $(x _{\alpha}, X _{\alpha})$ by the SDEs 
(\ref{eq:SDEx}) and (\ref{eq:SDEsq}).
This scheme is denoted by MQSQ-SP.
The propagation by 
Eqs. (\ref{eq:TDSE})--(\ref{eq:barbarV})
conserves the total energy expectation 
\(
\langle E \rangle = \langle \psi | \hat{H} | \psi \rangle
= \langle \varphi | \hat{T}_x | \varphi \rangle + \tilde{H}
\),
but the conservation is lost 
once the SPs are introduced via
Eqs. (\ref{eq:TDSESP})--(\ref{eq:HextSP}).
In this regard, parallel investigation of these two schemes will be useful
in practical studies.

Before proceeding to the numerical application, 
we note the relation between the MQSQ-SP and the MQC schemes.
By the classical point particle approximation for the heavy $X$ part
\begin{equation}
\left| \chi _{\Gamma} (X) \right|^2
\to
\delta (X - Q _t) ,
\end{equation}
we find
\(
V ( x ; \Gamma _t )
\to
v (x, Q _t)
\)
in Eq. (\ref{eq:barV}),
and then $U_{\gamma} (\Gamma)$ in Eq. (\ref{eq:barbarV}) is replaced by
\begin{equation}
U_{\gamma} ( Q )
=
\int dx \left| \varphi _{\gamma Q} ( x ) \right|^2 v(x, Q) .
\end{equation}
By introducing the Bohmian particles for the quantal $x$ part
and replacing the potential energy $U$ 
by the bare $v(x,X)$,
a MQCB scheme analogous to the previous ones \cite{Prezhdo01,Gindensperger00} is obtained.
We also note that the present MQSQ-SP has some similarity to
the time-dependent quantum Monte-Carlo method \cite{Christov07}. 
The apparent and most significant difference is in the deployment of SQ WP.

As a numerical demonstration, we study the same model as in Refs. \cite{Prezhdo01,Sholl98}
for gaseous O$_2$ collision to a Pt surface,
a prototype in which the ordinary MQC mean-field (MQC-MF) method fails
to describe the temporal splitting of the wave function to trapped and scattered parts.
The potential function is given by
%\begin{equation}
\begin{align}
v(x, X) = &
\frac{1}{2}M \Omega^2 X^2
+ a \left[ {\rm e}^{-2b(x-c)} - 2{\rm e}^{-b(x-c)} \right]
\nonumber \\
 & + A \; {\rm e}^{-B(x-X)} .
%\end{equation}
\end{align}
The first term is a harmonic binding potential of the heavy particle $X$ to the surface,
the second term is a Morse potential for the interaction between the light particle $x$ and the surface,
and the third term is a repulsive interaction between the particles.
For this $v(x,X)$, the $V$ of Eq. (\ref{eq:barV}) is derived as
%\begin{equation}
\begin{align}
V(x; \Gamma) = &
\frac{1}{2}M \Omega^2 \left( X^2 + \rho^2 \right)
+ a \left[ {\rm e}^{-2b(x-c)} - 2{\rm e}^{-b(x-c)} \right]
\nonumber \\
 & + A \; {\rm e}^{-B(x-X) + (B \rho)^2/2} .
%\end{equation}
\end{align}

\begin{figure}[t]
\begin{center}
\includegraphics[width=0.45\textwidth]{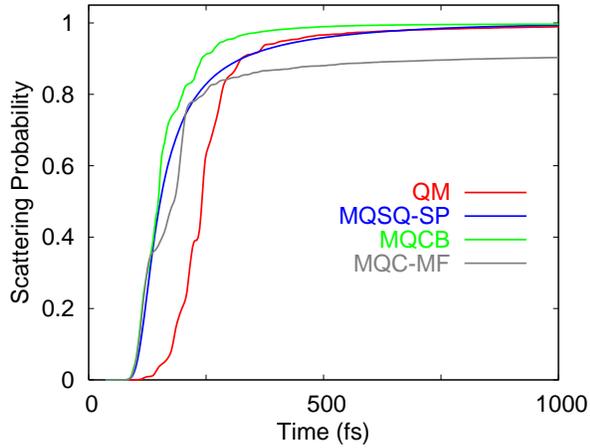}
\end{center}
\caption{
(Color online) 
Scattering probability from the methods of
full quantum mechanics (QM),
mixed quantal-semiquantal with stochastic particles (MQSQ-SP),
mixed quantum-classical with Bohmian particles (MQCB),
and
mixed quantum-classical mean-field approximation (MQC-MF).
}
\end{figure}

The initial wave function at $t=0$ is set as
a product of the harmonic ground-state wave function for $X$ and
a Gaussian WP for $x$ centered at $x=x_0$ with a width $\gamma$ and
the momentum $k_0$,
\begin{equation}
\psi(x, X, 0) 
= N 
\exp \left[- \frac{M \Omega X^2}{2 \hbar}  \right]
\exp \left[ - \frac{(x-x_0)^2}{\gamma^2} + \frac{i k_0 x}{\hbar} \right] ,
\end{equation}
in which 
$N = \left( 2 M \Omega / \pi^2 \hbar \gamma^2 \right)^{1/4}$.
The initial momentum $k_0$ is specified by the energy $E_0$ via
$k_0 = - \sqrt{2mE_0}$.
We have taken 
the numerical parameters from Ref. \cite{Prezhdo01}:
$m = 1$ amu, $M = 10$ amu,
$\Omega = 5 \times 10^{14} \textrm{ s}^{-1}$,
$A = 10^{4}$ kJ/mol,
$B = 4.25 \textrm{ \AA}^{-1}$,
$a = 700$ kJ/mol,
$b = 5.0 \textrm{ \AA}^{-1}$,
$c = 0.7$ \AA, 
$x_0 = 6.0$ \AA, 
and
$\gamma = 0.5$ \AA.
The quantum mechanical (QM) wave functions were propagated using 
Cayley's hybrid scheme with real-space grids \cite{Watanabe00}.
Convergence and unitarity of the propagation were
confirmed with 
the grid lengths $\Delta x = 0.0178$ \AA, $\Delta X = 0.0159$ \AA,
and the time step $\Delta t = 0.0124$ fs. 
The trajectories of ($Q, P$) and ($\rho, \Pi$) conjugate pairs
were propagated by
Suzuki's symplectic fourth-order scheme \cite{Suzuki90}.
The transmission-free absorbing potential \cite{GonzalezLezana04} was applied 
to the scattered wave function along $x$.
The results presented are with the absorbing potential set at 81 \AA\ $< x <$ 91 \AA,
although converged results were obtained with 45 \AA\ $< x <$ 51 \AA.
For the number of SP pairs, convergence was found with $N_{\rm sp} = 2000$.
The same number of Bohmian particles were used in the MQCB calculation.

Figure 1 presents the scattering probability defined by
\begin{equation}
P_{\rm s} (t) = 
\int_{x_{\rm s}}^{\infty} dx
\int_{-\infty}^{\infty} dX
\; | \psi (x, X, t) |^2 ,
\end{equation}
with $x_{\rm s} =$ 5.8 \AA\ \cite{Sholl98}.
The MQSQ-SP reproduces the correct asymptotic behavior, 
in contrast to the MQC-MF
and with improvement over the MQCB. 
However, the description of
delayed initial increase of QM $P_{\rm s}(t)$, due to the temporal resonance trapping
by the heavy particle excitation \cite{Sholl98}, was still incomplete.
In this regard, an intriguing further test 
would be to introduce dissipation to the heavy part.

\begin{figure}[t]
\begin{center}
\includegraphics[width=0.45\textwidth]{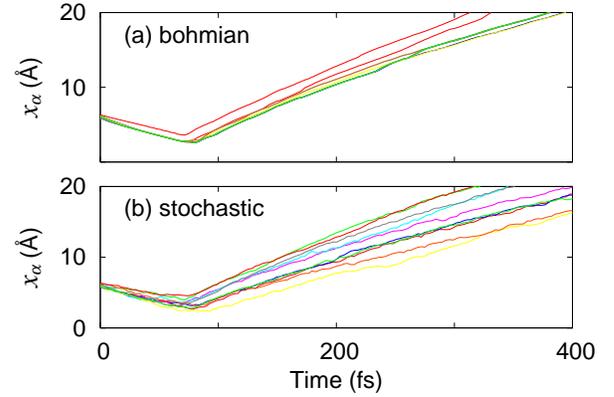}
\end{center}
\caption{
(Color online) 
Sample (ten) trajectories of 
(a) Bohmian particles in MQCB and 
(b) stochastic particles in MQSQ-SP.
}
\end{figure}

In an aim to understand the improved description,
we plot in Fig. 2 sample
trajectories of stochastic and Bohmian particles.
The difference basically emerges from the osmotic term $\nabla R / m$ 
and the stochastic term $\sqrt{\hbar/m} \; dW$ in Eq. (\ref{eq:SDEx});
the Bohmian dynamics do not involve them but
only the current velocity $dx_t = \nabla S / m$.
This provides an understanding of the more ballistic trajectories of Bohmian in Fig. 2a.
However, further analysis revealed that 
the use of SPs alone does not account for the difference,
because
a combination of MQC and SP resulted in $P_{\rm s}(t)$
almost identical 
to that from MQCB,
which indicates that 
the combination of MQSQ and SP is essential for the result in Fig. 1.

In summary, we have formulated a MQSQ theory
with a SP description of the interparticle correlation,
and examined it numerically for a prototype model 
involving
wave function splitting.
Despit its simplicity, the results were encouraging, 
although a need for refining the description of interparticle
correlation was still evident.
We also note that the model employs 
for the heavy part a harmonic potential on which
classical mechanics is patently appropriate.
More stringent tests should clarify the nature of 
the present MQSQ scheme.
Particularly interesting would be the cases in which 
the quantum mechanical aspects of the heavy part play some role, for instance,
in the zero-point energy leakage \cite{Habershon09}.

Finally, we note that the SQ WP of Eq. (\ref{eq:SQWP}) can be regarded as 
a coherent state basis 
for the path-integral formulation of quantum propagator \cite{Kuratsuji81}.
We have recently demonstrated that the initial value representation 
of the propagator in combination with the SQ WP 
is applicable \cite{Ando14}.
This will 
provide more flexible description of the wave function by the proper inclusion of
quantum phase. 
Its integration with the present MQSQ formulation is a direction in which to proceed.

The author acknowledges support from KAKENHI Nos. 22550012 and 26620007.

\end{document}